\documentclass[aps,pre,twocolumn,superscriptaddress,showpacs,showkeys ]{revtex4-1}

\usepackage{graphicx}                   
\usepackage{hyperref}                   
\usepackage{amsmath,amssymb}            
\usepackage{epstopdf}
\usepackage{subcaption}					

\usepackage{color}
\definecolor{orange}{rgb}{1,0.5,0}
\definecolor{brown}{rgb}{0.65, 0.16, 0.16}
\definecolor{phlox}{rgb}{0.87, 0.0, 1.0}

\begin{document}

\title{Schramm-Loewner evolution in the random scatterer Henon-percolation landscapes}

\author{M. N. Najafi}
\affiliation{Department of Physics, University of Mohaghegh Ardabili, P.O. Box 179, Ardabil, Iran}
\email{morteza.nattagh@gmail.com}

\author{S. Tizdast}
\affiliation{Department of Physics, University of Mohaghegh Ardabili, P.O. Box 179, Ardabil, Iran}
\email{Susan.tizdast@gmail.com }

\author{J. Cheraghalizadeh}
\affiliation{Department of Physics, University of Mohaghegh Ardabili, P.O. Box 179, Ardabil, Iran}
\email{jafarcheraghalizadeh@gmail.com }

\begin{abstract}
The Shcramm-Loewner evolution (SLE) is a correlated exploration process, in which for the chordal set up, the tip of the trace evolves in a self-avoiding manner towards the infinity. The resulting curves are named SLE$_{\kappa}$, emphasizing that the process is controlled by one parameter $\kappa$ which classifies the conformal invariant random curves. This process when experiences some environmental imperfections, or equivalently some scattering random points (which can be absorbing or repelling) results to some other effective scale-invariant curves, which are described by the other effective fractal dimensions and equivalently the other effective diffusivity parameters $\kappa_{\text{effective}}$. In this paper we use the classical Henon map to generate scattering (absorbing/repelling) points over the lattice in a random way, that realizes the percolation lattice with which the SLE trace interact. We find some meaningful power-law changes of the fractal dimension (and also the effective diffusivity parameter) in terms of the strength of the Henon coupling, namely the $z$ parameter. For this, we have tested the fractal dimension of the curves as well as the left passage probability. Our observations are in support of the fact that this deviation (or equivalently non-zero $z$s) breaks the conformal symmetry of the curves. Also the effective fractal dimension of the curves vary with the second power of $z$, i.e. $D_F(z)-D_F(z=0)\sim z^2$. 
\end{abstract}


\maketitle

\section{Introduction}

Schramm-Loewner evolution (SLE) is served as a powerful tool in classifying two-dimensional (2D) critical statistical systems in one parameter classes~\cite{Schramm,Cardy}. In contrast to the conformal field theory (CFT) which deals with the local fields~\cite{francesco2012conformal}, it analyzes the global quantities of the model in hand and maps the problem into a dynamical exploration process~\cite{Cardy,NajafiPRE1}. This is done by using a stochastic equation which governs the uniformzing conformal maps, by which an important parameter, namely the diffusivity parameter $\kappa$ is extractaed. This dynamical equation is also referred to as the Loewner process~\cite{Schramm}. The diffusivity parameter is referred to as the parameter of the universality class of 2D conformal systems~\cite{BauBer}.\\
The statistical analysis of such systems becomes possible by mapping of the extended interfaces of the 2D conformal systems (which can be the domainwalls of the separated phases, or something else, which is commonly resultant of the \textit{snapshots} of a dynamical systems), to the exploration (Loewner) process. This method has been proved to be very promissing in recognizing and classifying the well-known 2D conformal systems as well as the less-known ones. The examples for the well-known models are the Ising model~\cite{Najafi2015Observation}, the Coulomb gases~\cite{Nienhuis}, the loop erased random walks~\cite{BerBau}, and the self-avoiding walks~\cite{SchrammSAW}. A good review on the subject can be found in~\cite{Cardy}. The most important examples for the less-known models are also the BTW sandpile model~\cite{Najafi2012observation,Najafi2013water,Najafi2016Bak}, the watersheds~\cite{Daryaei2012Watersheds}, the graphene system~\cite{Herrmann,Najafi2017Scale}, the normal state of the YBCO superconducting planes~\cite{Najafi2016Universality}, the random field Ising model~\cite{randomfield}, the Ising model on the percolation systems~\cite{Najafi2016Monte}, the $(2+1)$-dimensional growing surfaces~\cite{Saberigrowing}, the Darcy model of fluid propagation in the porous media~\cite{Najafi2015Geometrical} and the three-state Potts model~\cite{threepotts}. Many aspects of this model are known, ranging from its Fokker-Planck equation~\cite{NajafiPRE2}, to its relation to the other models like the Coulomb gases~\cite{Cardy}, and also many statistical features (SLE predictions) of the the random curves are kown, such as the left passage probability~\cite{NajafiPRE1}, the fractal dimension, and the Cardy's crossing probability~\cite{Cardy}. \\
The correlated dynamical paths in the SLE theory, whatever are they describing, are defined in the upper half plane in the chordal set up. This area is regular, and each point of the upper half plane has the chance to be visited through an exploration process. A question may arise concerning making of the space partially imperfect by distributing some \textit{forbidden} or \textit{inactive} areas over the host system, i.e. the upper half plane. In the other words, what happens if the dynamical traces of SLE (which are described by the Loewner process in the regular system) are not allowed to enter some quenched random regions? This is the main aim of the present paper.\\
Actually the answer of this question is not simple from both practical and theoretical point of view, since there are some problems in its realization. More precisely, the SLE paths are built by means of some stochastic Lengevin equations, that move simultaneously all points of the trace (up to some definite time), which yields the SLE trace at final. Therefore, we cannot force the points not to enter the forbidden regions. A possible way out of this problem is to couple the SLE Langevin equation with some other equations which absorb (repel) the traces of the SLE towards (from) the allowed (forbidden) sites. There are many ways to do so, one of which is the using of the two-dimensional dynamical maps, that should be designed carefully by demanding some points of the space are stable fixed points and someones are unstable fixed points of the map, which is done in the present paper. The Henon map is a proper candidate for which the classical fixed points can be easily designed to contain absorbent/repellent fixed points. By choosing the configuration of absorbent/repellent (scattering) sites completely randomly, we have actually coupled the SLE theory with the percolation theory. We have also put a tunable parameter $z$ in the analysis that tunes the strength of the Henon scatterer (absorbent/repellent) sites. \\
We have considered four $\kappa$ values in the dilute SLE phase, i.e. $\kappa=2,\frac{8}{3},3$ and $4$, and tested the effect of the Henon-percolation lattice on them which is tuned by the parameter $z$. In each case we have extracted the effective diffusivity parameter $\kappa_{\text{effective}}$ of the resultant curves. To do so, we have used two parallel tests: the fractal dimension of the curves and the left passage probability test. We have interestingly observed some power-law behaviors in terms of $z$. We see in the followings that the conformal symmetry of the (initially conformal) random traces breaks down.\\
The paper has been organized as follows: in the following section we introduce shortly the SLE theory. Section~\ref{SEC:model} has been devoted to the definition of the problem. In that section we derive the main equations of our claim. After describing the numerical methods of the paper in SEC.~\ref{SEC:num}, we will present our main findings in SEC.~\ref{SEC:results}. We will close the paper by a conclusion.

\section{Schramm-Loewrner Evolution}\label{SEC:SLE}

According to the SLE theory one can describe the geometrical objects (which may be interfaces) of a 2D critical model via a growth process and classify them into one parameter ($\kappa$) classes \cite{Cardy}. From a simple relation between the central charge $c$ in CFT and the diffusivity parameter $\kappa$ in SLE, namely $c=\frac{(6-\kappa)(3\kappa-8)}{2\kappa}$, one can find the corresponding CFT \cite{Cardy,NajafiPRE1,NajafiPRE2}, and consequently the universality class is obtained. Chordal SLE$_{\kappa}$ is a growth process defined via the conformal maps, $g_{t}(z)$, which are solutions of the Loewner's equation:
\begin{equation}
\partial_{t}g_{t}(z)=\frac{2}{g_{t}(z)-\xi_{t}}
\end{equation}
where the initial condition is $g_{t}(z)=z$ and $\xi_{t}$ (the driving function) is a continuous real valued function which is shown to be proportional to the one dimensional Brownian motion ($\xi_t=\sqrt{\kappa}B_t$) if the curves have two properties: conformal invariance and the domain Markov property. \\
One of the most important measures of the SLE theory is the left passage probability (LPP), that is the probability that an arbitrary point $z=x+iy$ falls to the right of a chordal curve, i.e. the curve passes from the left of $z$. It has been shown that~\cite{NajafiPRE1}:
\begin{equation}
\begin{split}
\text{LPP}(\theta)&=\frac{1}{2}+\frac{\Gamma(4/\kappa)}{\sqrt{\pi}\Gamma((8-\kappa)/2\kappa)}{_2}F_1\left[\frac{1}{2},\frac{4}{\kappa},\frac{3}{2},-\left( \cot(\theta)\right)^2 \right]\\
&\times \cot(\theta)
\end{split}
\label{Eq:LPP}
\end{equation}
in which $\cot(\theta)\equiv \frac{y}{x}$, and ${_2}F_1$ is a hypergeometric function. The above equation declares that if the curve is conformal, then LPP should be independent of $r\equiv\sqrt{x^2+y^2}$. The fractal dimension of the stochastic curve can also be served as another important measure which is related to $\kappa$ via the relation $D_F=1+\frac{\kappa}{8}$~\cite{Cardy}. In the present paper we regularly use these equations to extract the effective diffusivity parameter which is defined in the following section.

\begin{figure}
	\centerline{\includegraphics[scale=.35]{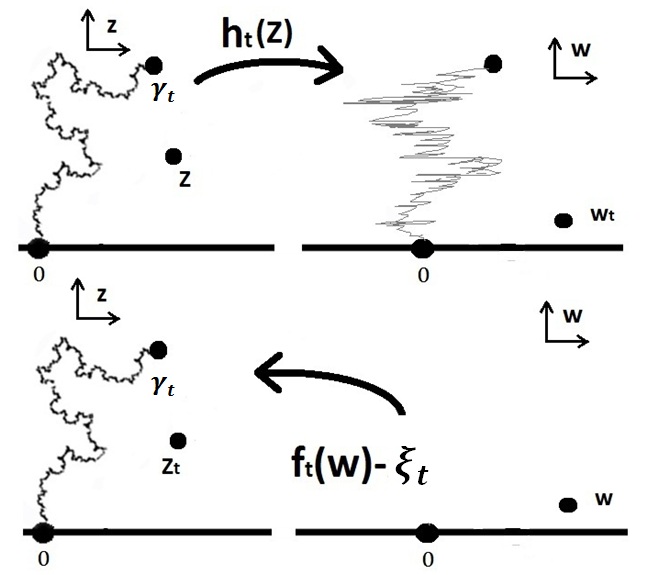}}
	\caption{The procedure of the inverse SLE map, resulting to retrieval of the SLE path. Upper: The path of the tip of the trace by repeatedly applying the SLE uniformizing map. Lower: The inverse procedure using the $f_t$ mapping.}
	\label{fig:FPE}
\end{figure}

Now we are going to describe the Langevin equation for the chordal SLE trace. For details see~\cite{Gruzberg}. Let us first define the shifted conformal map $h_t(z)$ as follows:
\begin{equation}
h_t(z)\equiv g_t(z)-\zeta_t
\end{equation}
for which one can easily verify that
\begin{equation}
h_t^{-1}(w)\stackrel{d}{=}g_t^{-1}(w+\zeta_t).
\label{equivlence}
\end{equation}
In the above equation $\stackrel{d}{=}$ means the equality of the distributions of stochastic processes. The differential equation governing $h_t$ is:
\begin{equation}
\begin{split}
\partial_th_t(z)=\frac{2}{h_t(w)}-\partial_t\zeta_t
\end{split}
\label{h_eq}
\end{equation}
in which $h_0(z)=z$. One can retrieve the SLE trace by the relation $\gamma(t)=\lim_{\epsilon\downarrow{0}}g_{t}^{-1}(\zeta_{t}+i\epsilon)=\lim_{z\rightarrow \zeta_t}g_t^{-1}(z)$ and find the differential equation of the tip of SLE trace. The equation governing $g_t^{-1}$ is very difficult to solve. There is a way out of this problem, using the backward SLE equation. The backward SLE mapping $f_t(z)$ is defined as follows (note that $\zeta_t\stackrel{d}{=}\zeta_{-t}\stackrel{d}{=}-\zeta_t$) \cite{Gruzberg}:
\begin{equation}
\begin{split}
\partial_tf_t(w)=-\frac{2}{f_t(w)-\zeta_t}
\end{split}
\label{backw}
\end{equation}
It has been shown that the probability distribution of $f_t$ is the same as $g_t^{-1}$ \cite{Gruzberg}, i.e.
\begin{equation}
\begin{split}
f_t(w)-\zeta_t\stackrel{d}{=} g_t^{-1}(w+\zeta_t)\stackrel{d}{=}h_t^{-1}(w).
\end{split}
\label{equivlence2}
\end{equation}
The schematic representation of this equation has been sketched in FIG.~\ref{fig:FPE}. Therefore the tip of the SLE trace can be obtained by $\gamma_T=\lim_{w\rightarrow 0}g_T^{-1}(w+\zeta_T)=\lim_{w\rightarrow 0}h_T^{-1}(w)$. Now we can find the trajectory of the tip of the SLE trace using the backward equation (\ref{backw}) for $z_t=x_t+y_t$ which is (notice that Re$(\gamma_t)$ and Im$(\gamma_t)$ have the same joint distribution as $x_t$ and $y_t$)
\begin{equation}
\begin{split}
& \partial_tx_t=-\frac{2x_t}{x_t^2+y_t^2}-\partial_t\zeta_t\\
& \partial_ty_t=\frac{2y_t}{x_t^2+y_t^2}
\end{split}
\label{BKWComponents}
\end{equation}
conditioned to have the initial values $x_0=u$ and $y_0=v$ in which $w=u+iv$ is the initial point of the flow. This is the starting point of our analysis, i.e. the above equations provide the possibility of building numerically the real SLE traces, and are regularly used in this paper.

\section{exploration process in the landscape of random scatterers in the Henon-percolation lattice}\label{SEC:model}

\begin{figure}
	\centerline{\includegraphics[scale=.37]{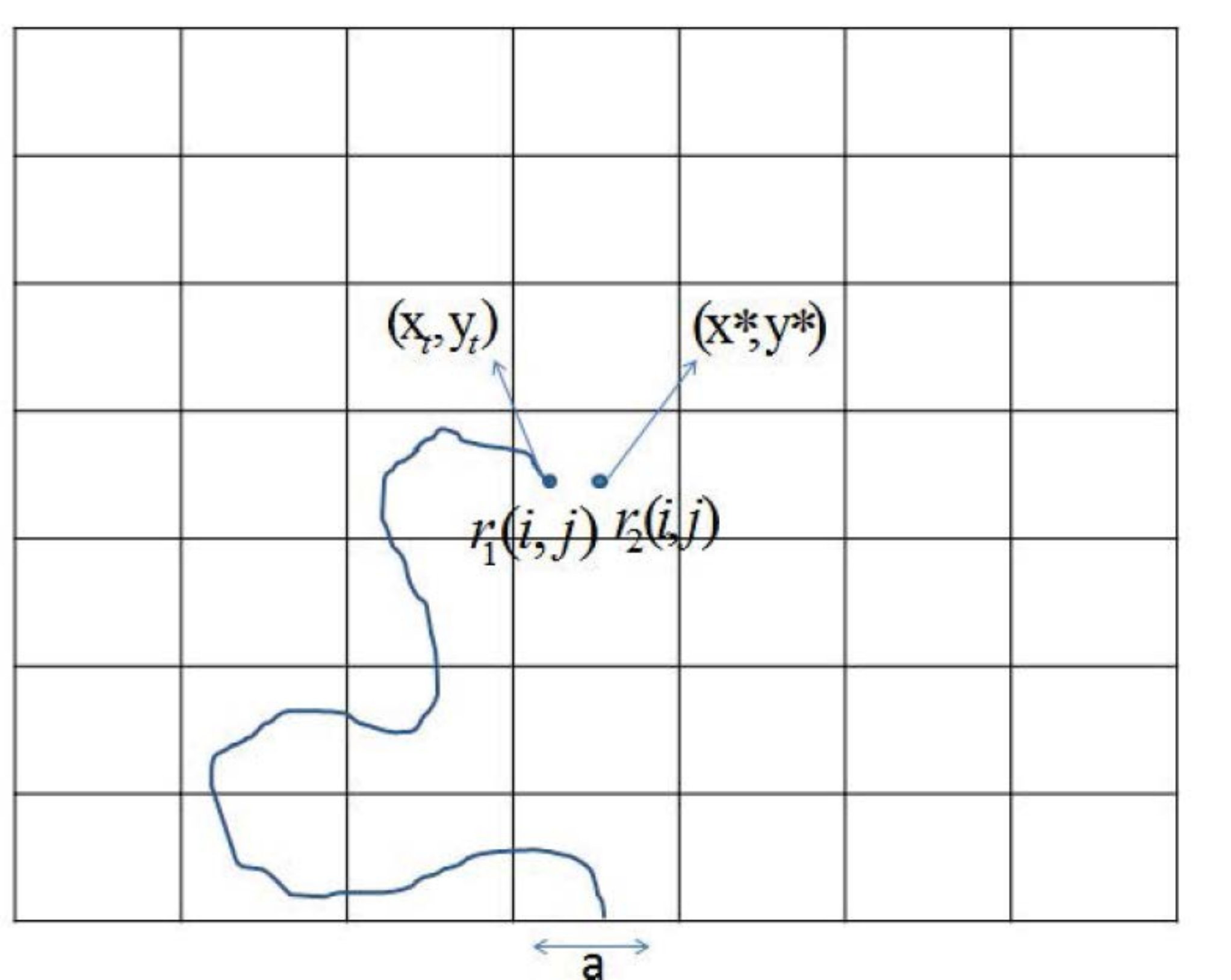}}
	\caption{The schematic arrangement of the problem of SLE trace correlated to the lattice interactive points. When the tip of the graph reaches a block, the point which is located right in the middle of the block plays the role of the unstable/stable fixed point which drives out/in the tip.}
	\label{fig:Scehmatic1}
\end{figure}
In this section we theoretically present a method to make SLE traces interactive with lattice sites, i.e. declare how they are put into a percolation media in which there are some centers of absorption and repulsion which interact with the SLE trace. This is possible via inserting some absorption/repulsion (scattering) centers in the dynamical equation of SLE trace. Let us go into details by rewriting the Langevin equation of non-interactive SLEs, i.e. Eq.~\ref{BKWComponents} as follows:
\begin{equation}\label{Langevin} 
\left\{ \begin{array}{l}
\dot x =  - \frac{{2x}}{{{x^2} + {y^2}}} - \sqrt \kappa  \dot B\\
\dot y = \frac{{2y}}{{{x^2} + {y^2}}}
\end{array} \right.
\end{equation}
As stated above, we intend to use the Henon map to generate the absorption/repulsion (scattering) centers over the lattice. The Henon map is normally defined as follows:
\begin{equation}
\left\{ \begin{array}{l}
x_{n + 1}-x_0 = (y_n-y_0) + 1 - r_1(x_n-x_0)^2\\
y_{n + 1}-y_0 = r_2(x_n-x_0)
\end{array} \right. ,
\end{equation}
in which $r_1$ and $r_2$ are the control parameters of the map, and $(x_0,y_0)$ is the origin or the reference point. In the continuum limit this equation transforms to:
\begin{equation}\label{Henon} 
\left\{ \begin{array}{l}
\dot x = (y-y_0) + 1 - {r_1}(x-x_0)^2 - (x-x_0)\\
\dot y = {r_2}(x-x_0) - (y-y_0)
\end{array} \right. .
\end{equation}
The fixed points of the Henon map can be readily found by demanding $ x_{n + 1}^ *  = x_n^ *  $ and $ y_{n + 1}^ *  = y_n^ *  $, which are found to be:
\begin{equation}
\left\{ \begin{array}{l}
x_n^ *-x_0  = \frac{{{r_2} - 1}}{{2{r_1}}} \mp \frac{1}{{2{r_1}}}\sqrt {{{\left( {{r_2} - 1} \right)}^2} + 4{r_1}} \\
y_n^ *-y_0  = {r_2}x_n^ * 
\end{array} \right.
\end{equation}
or equivalently:
\begin{equation}
\left\{ \begin{array}{l}
{x_0} = {x^ * } - \left( {\frac{{{r_2} - 1}}{{2{r_1}}} \mp \frac{1}{{2{r_1}}}\sqrt {{{\left( {{r_2} - 1} \right)}^2} + 4{r_1}} } \right)\\
{y_0} = {y^ * } - {r_2}x_2^ * 
\end{array} \right.
\label{Eq:x_0y_0}
\end{equation}
We can set $x^*$ and $y^*$ to be an arbitrary point in space and also tune $r_1$ and $r_2$ to make the point unstable (repulsive) or stable (absorptive). Then the corresponding Henon process, when is started from the true reference point $(x_0,y_0)$, is driven in/out (depending to the chosen values of $r_1$ and $r_2$) to/from the point $(x^*,y^*)$.\\
Now we are in the position to consider the main problem of the paper, i.e. we combine the SLE dynamical equation with the Henon map to make the SLE trace interactive with the lattice points. The most direct way is to insert the right-hand-side expressions of Eq.~\ref{Henon} into Eq.~\ref{Langevin}. By doing so we obtain:
\begin{equation}\label{d} 
\left\{ \begin{array}{l}
\dot x =  - \frac{{2x}}{{{x^2} + {y^2}}} - \sqrt \kappa  \dot B +z \left[ y-y_0 + 1 - {r_1}{(x-x_0)^2} - (x-x_0)\right] \\
\dot y = \frac{{2y}}{{{x^2} + {y^2}}} +z \left[ {r_2}(x-x_0) - (y-y_0)\right] 
\end{array} \right.
\end{equation}
or in the discrete case ($\delta x_n\equiv x_n-x_0$, and $\delta y_n\equiv y_n-y_0$)
\begin{equation}
\left\{ \begin{array}{l}
{x_{n + 1}} = {x_n} - \left( \frac{{2{x_n}}}{{x_n^2 + y_n^2}} + z \left[ {r_1}\delta x_n^2 + \delta x_n - {\delta y_n} - 1\right]  \right)\delta t\\
- \sqrt \kappa  \delta {B_n}\\
{y_{n + 1}} = {y_n} + \left( \frac{{2{y_n}}}{{x_n^2 + y_n^2}} + z \left[ {r_2}{\delta x_n} - \delta y_n\right]  \right)\delta t
\end{array} \right.
\label{Eq:SLE-Henon}
\end{equation}
in which the $z$ parameter controls the strength, or contribution of the Henon map. In this process the SLE traces are partially driven in (out) to (from) the point $(x^*,y^*)$. This enables us to set $x^*$ and $y^*$ a point of interest. \\
Especially we can choose $(x^*,y^*)$ to be the center of the squares of a two-dimensional square lattice. In this case $x^*= x_{\text{mid}} = a\left( {n + 0.5} \right) $ and $y^*=y_{\text{mid}} = a\left( {m + 0.5} \right)$, where $(n\equiv\left[\frac{x_n}{a}\right]+1,m\equiv\left[\frac{y_n}{a}\right]+1)$ is the coordinate of the selected square in the lattice, and $a$ is the lattice constant.\\
The general dynamical process is as follows: Before the process is started, we determine the fields $r_1$ and $r_2$ over the lattice, i.e. $ \left\lbrace r_1(n,m)\right\rbrace_{n,m=1}^{L}$ and $\left\lbrace r_2(n,m)\right\rbrace_{n,m=1}^{L}$. Accordingly, each block of the lattice become repulsive or attractive region for the SLE trace. The SLE trace evolves according to the Eq.~\ref{Eq:SLE-Henon}. When the trace enters a block $(n,m)$ in the lattice, $(x^*,y^*)$ is set to the middle of the block, and the corresponding $(x_0,y_0)$ is calculated via the Eq.~\ref{Eq:x_0y_0}, and is inserted to the dynamical equation~\ref{Eq:SLE-Henon}. This has been schematically shown in Fig.~\ref{fig:CurveSamples}, in which the tip of the SLE trace as well as the middle of the block have been sketched. $r_1$ and $r_2$ have been set to their values in the classical Henon map, with some uncorrelated fluctuations, i.e. $r_1(n,m) = 1.4 + C(\eta-0.5)$ and $r_2(n,m) = 0.3+ C(\eta-0.5)$, in which $\eta$ is some random number with uniform distribution in the $[0,1]$ interval, and $C$ is the strength of the randomness that is set to $0.2$ in this paper.

\section{The numerical details}\label{SEC:num}

\begin{figure*}
	\centerline{\includegraphics[scale=.60]{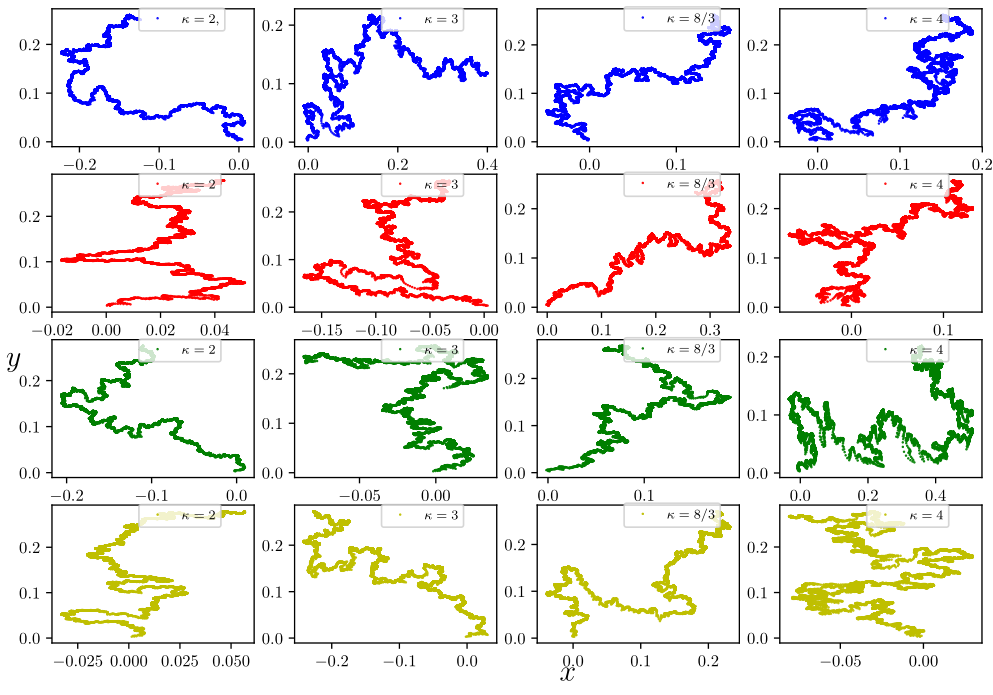}}
	\caption{The real SLE trace samples. In each Column the input $\kappa$ (the employed $\kappa$ for the SLE to by interact with Henon-percolation lattice) in fix, but $z$ varies from zero (top) to $8$ (bottom), i.e. $z=0,2,4$ and $8$ from top to bottom.}
	\label{fig:CurveSamples}
\end{figure*}

In this section we describe the numerical details and the ways that the critical exponents have been extracted. To generate the samples we have used the SLE-Henon equation~\ref{Eq:SLE-Henon}, and the random curves have started from the origin. When the trace of SLE reaches one block of the lattice, the only attractive/repulsive (scattering) point that acts on it is located at the center of that block. The size of blocks has been set to $a=10^{-3}$, and the time step of the SLE map has been chosen to be $\delta t=10^{-5}$, and the SLE traces have been allowed to grow over $3\times 10^{4}$ steps, i.e. the curves contain $3 \times 10^4$ points. The effect of Henon map is controlled by the $z$ parameter. \\
We have run the program for four different $\kappa$s: $\kappa=2,\frac{8}{3},3$ and $4$. We have also considered various rates of $z$ to control the effects of the Henon-percolation lattice: $z=0,1,2,4,8,32$ and $64$. The size of the system can be arbitrarily large, but for $N=3\times 10^4$ point curves, $L=10$ (corresponding to a lattice with linear size $L_x=\frac{L}{a}=10^4$) had been sufficient. Some samples of this problem have been shown in Fig.~\ref{fig:CurveSamples} in which $\kappa$ refers to the bare diffusivity parameter from which we start, that is constant over each column. In each column, $z$ varies from zero to higher values, i.e. $z=0,2,4$ and $8$ from top to bottom. 
\begin{figure*}
	\centering
	\begin{subfigure}{0.35\textwidth}\includegraphics[width=\textwidth]{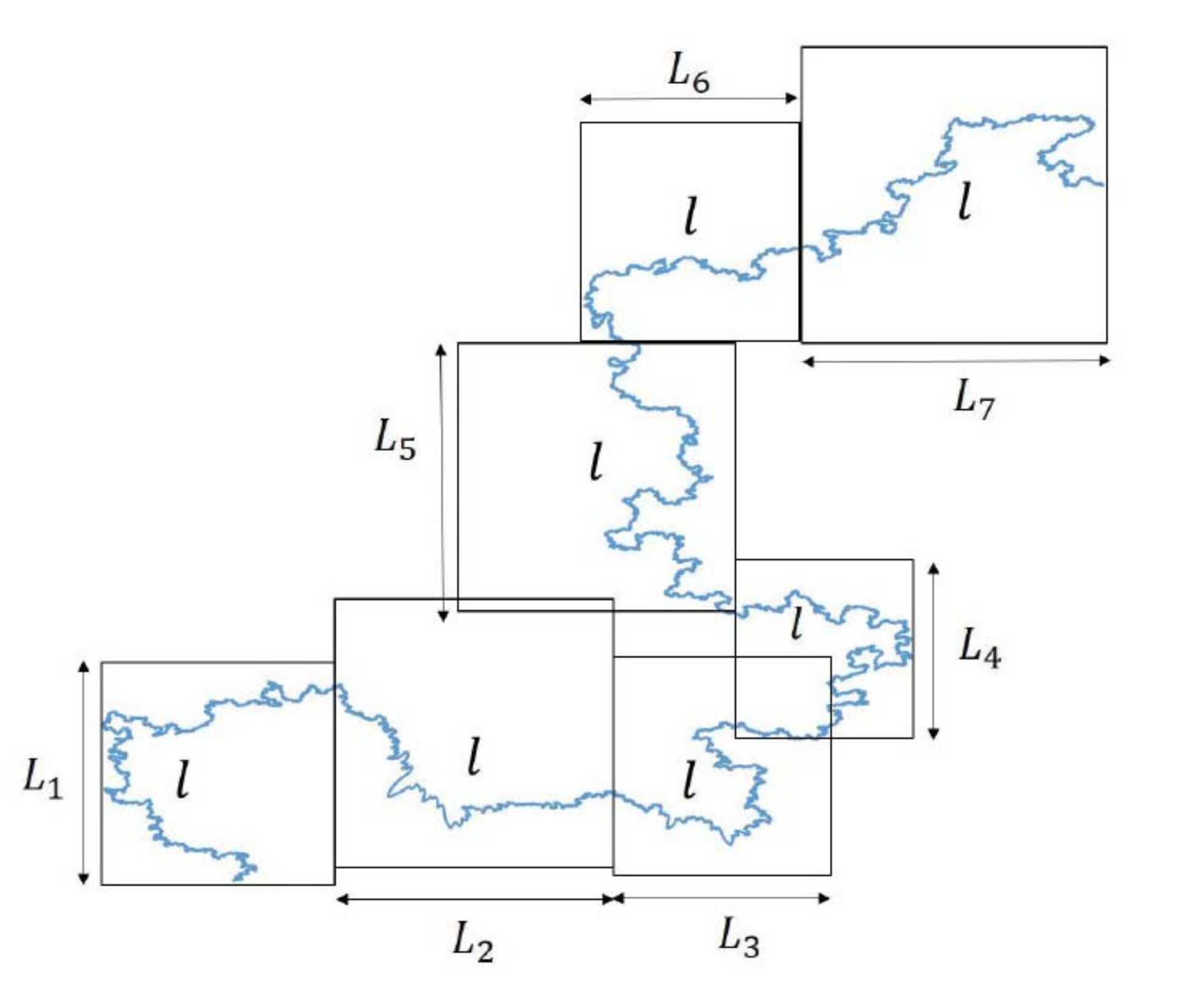}
		\caption{}
		\label{fig:Scehmatic2}
	\end{subfigure}
	\begin{subfigure}{0.35\textwidth}\includegraphics[width=\textwidth]{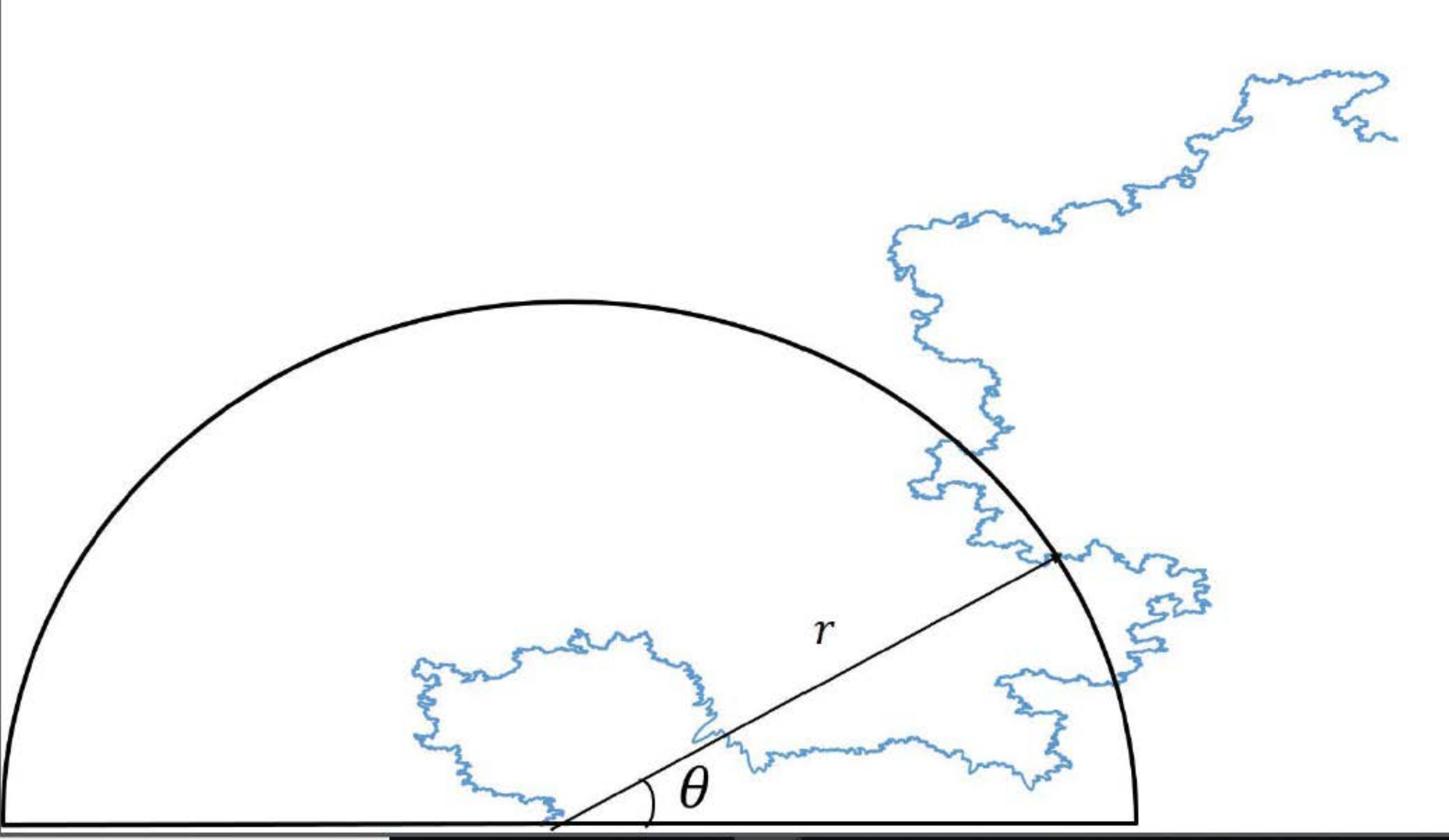}
		\caption{}
		\label{fig:Scehmatic3}
	\end{subfigure}
	\caption{(Color online) The procedure of calculating (a) the fractal dimension, and (b) the LPP function of the SLE curves.}
	\label{fig:Scehmatic23}
\end{figure*}
In the analysis of the curves, we have used two parallel tests to extract the presumable effective diffusivity parameter $\kappa_{\text{effective}}$, namely left passage probability (LPP) and the fractal dimension of the SLE traces. The numerical set up of these quantities have been shown in Figs.~\ref{fig:Scehmatic2} and~\ref{fig:Scehmatic3}. As is evident in this figure, the box-counting method has been used to find the fractal dimension (in which each box contains the same length of the curve). Also the LPP can be evaluated by varying the $\theta$ parameter from zero to the maximum value, which is $\pi$. If the curves are conformal invariant, then the LPP should not depend on $r$ (which is the case for all supposed $z$s, except very large ones $z=32$ and $64$). 
\section{Results and Discussion}\label{SEC:results}

\begin{figure*}
	\centering
	\begin{subfigure}{0.40\textwidth}\includegraphics[width=\textwidth]{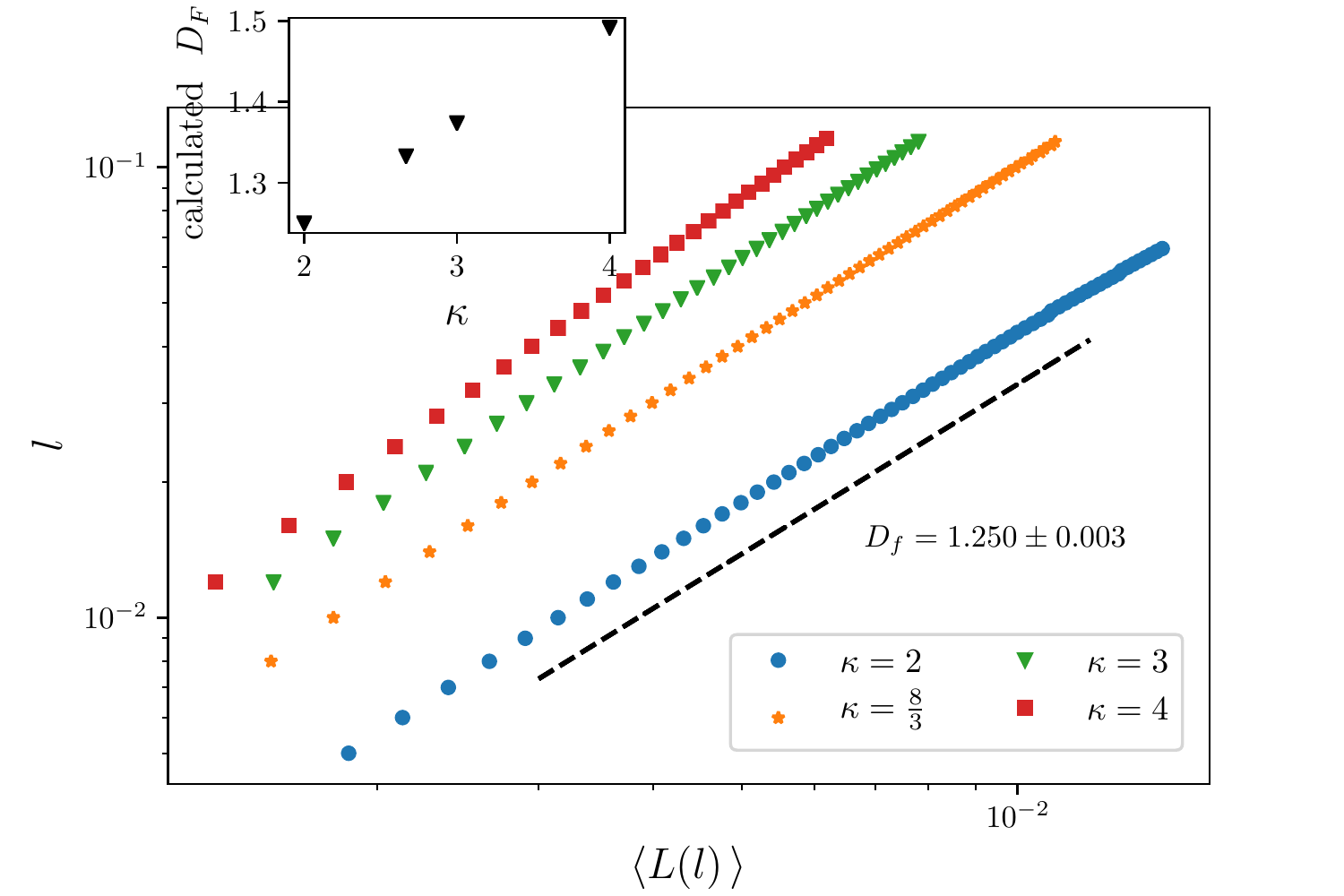}
		\caption{}
		\label{fig:D_f_z=0bc}
	\end{subfigure}
	\begin{subfigure}{0.45\textwidth}\includegraphics[width=\textwidth]{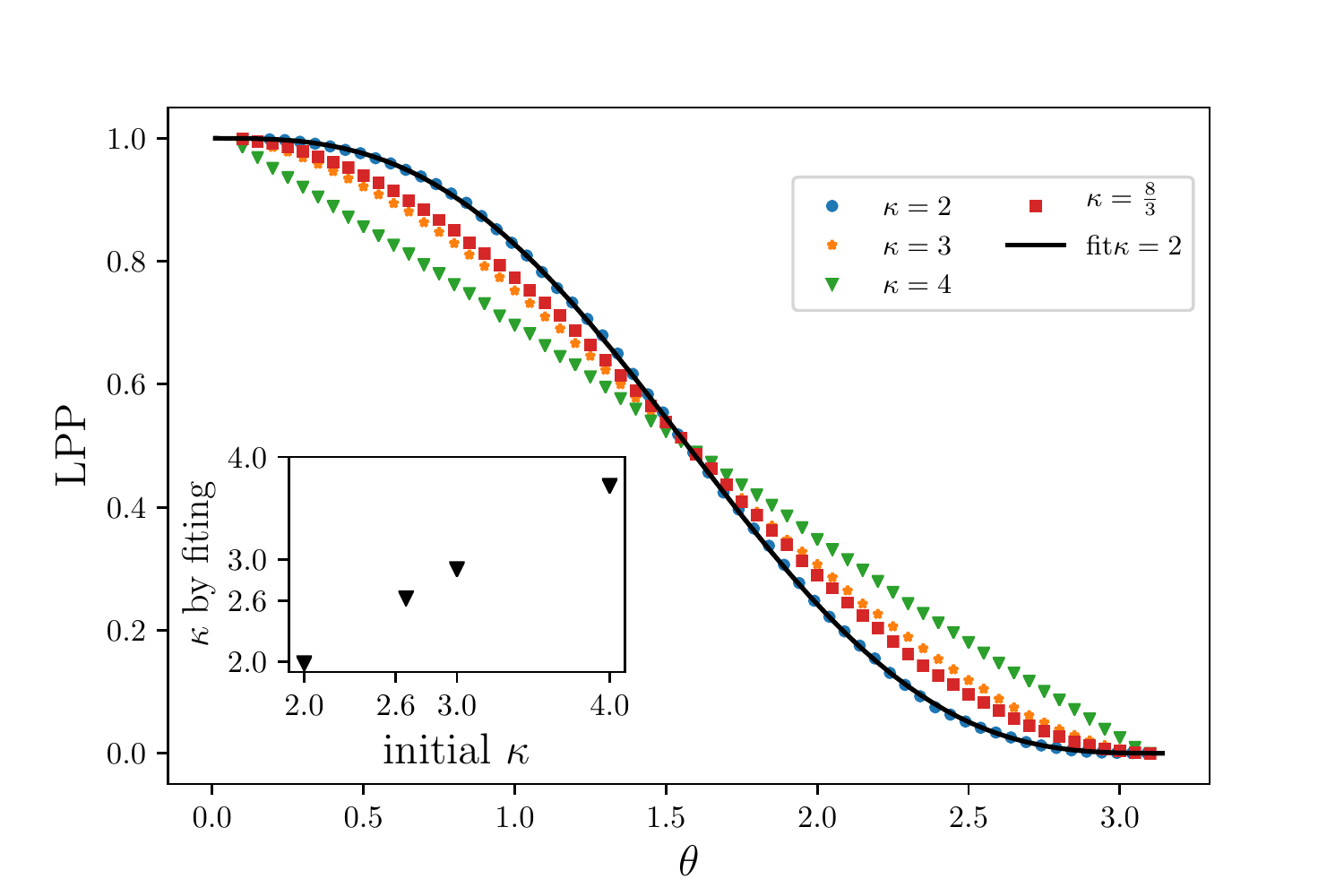}
		\caption{}
		\label{fig:z=0lpp}
	\end{subfigure}
	\caption{(Color online) The numerical results for the case $z=0$. (a) The fractal dimension, which is the slope of the $l-L$ graph in the log-log plot, and (b) the LPP function for the SLE curves. The fits have been shown for $\kappa=2$. The other fits (which have not been shown here) are as good as $\kappa=2$.}
	\label{fig:z=0}
\end{figure*}

In this section the results are presented. We should first examine our numerical procedure for generating SLE curves with arbitrary diffusivity parameters in the regular host, i.e. see if the numerical tests work for the case $z=0$. The equations~\ref{BKWComponents} should be employed for this purpose. If all the theoretical and numerical procedures are correct, then by an input $\kappa$ in Eq.~\ref{BKWComponents}, the output $\kappa$ should be the same. In the Fig.~\ref{fig:z=0} we have shown the two mentioned tests for $z=0$, and $\kappa=2,\frac{8}{3},3$ and $4$. In the Fig.~\ref{fig:D_f_z=0bc} the fractal dimension of the curves is the slope of the graph, which is completely in agreement of the theoretical expectations, i.e. $D_f=1+\frac{\kappa}{8}$ for all input $\kappa$s. The same is true for Fig.~\ref{fig:z=0lpp}, in which the fit has been shown for $\kappa=2$. Again a complete agreement with the theoretical value (Eq.~\ref{Eq:LPP}) is obtained. This is indeed the case for all examined $\kappa$s. All of these say that the methods that we have employed to extract the statistical quantities are reliable.

\begin{figure*}
	\centering
	\begin{subfigure}{0.40\textwidth}\includegraphics[width=\textwidth]{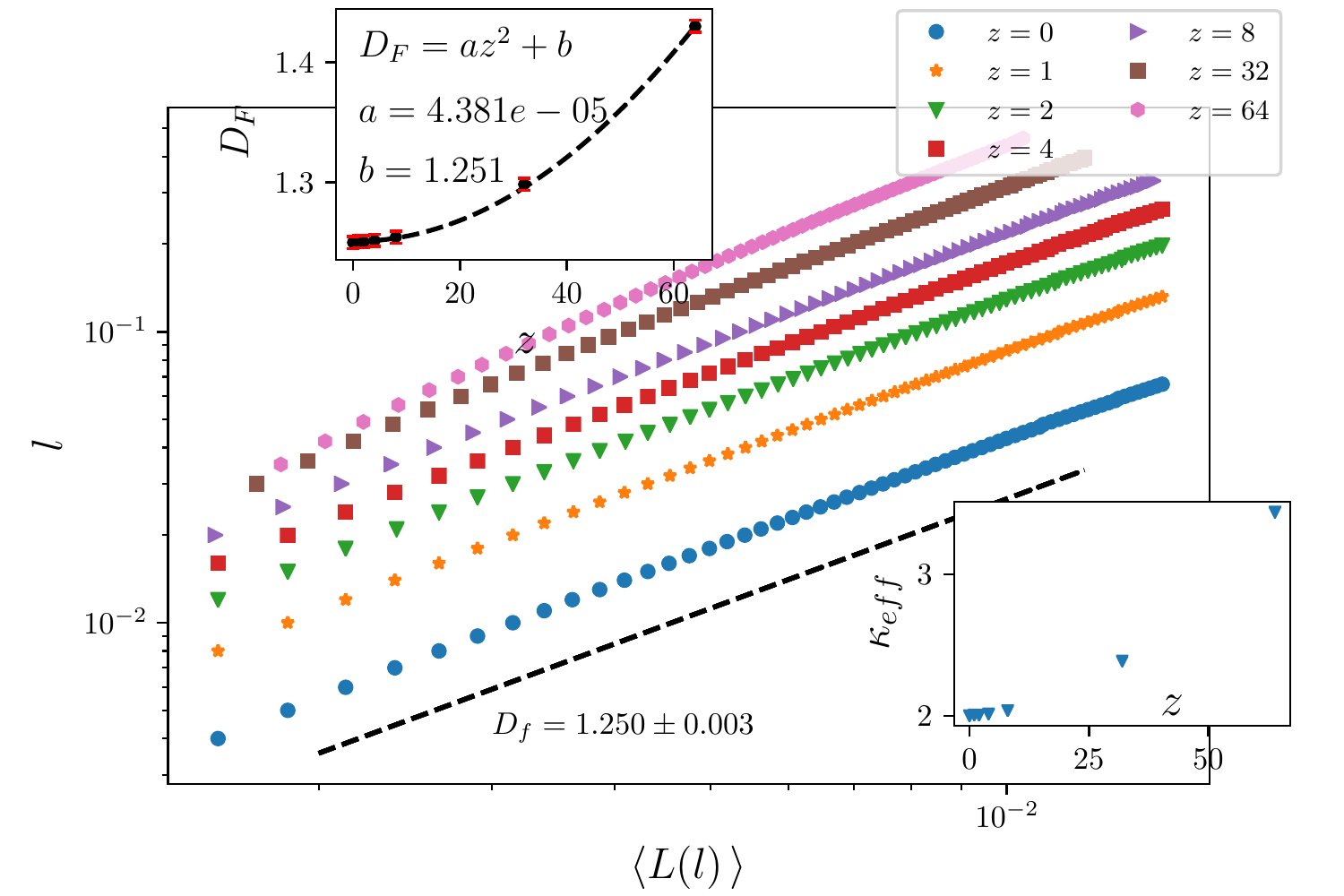}
		\caption{}
		\label{fig:D_f_k=2}
	\end{subfigure}
	\begin{subfigure}{0.45\textwidth}\includegraphics[width=\textwidth]{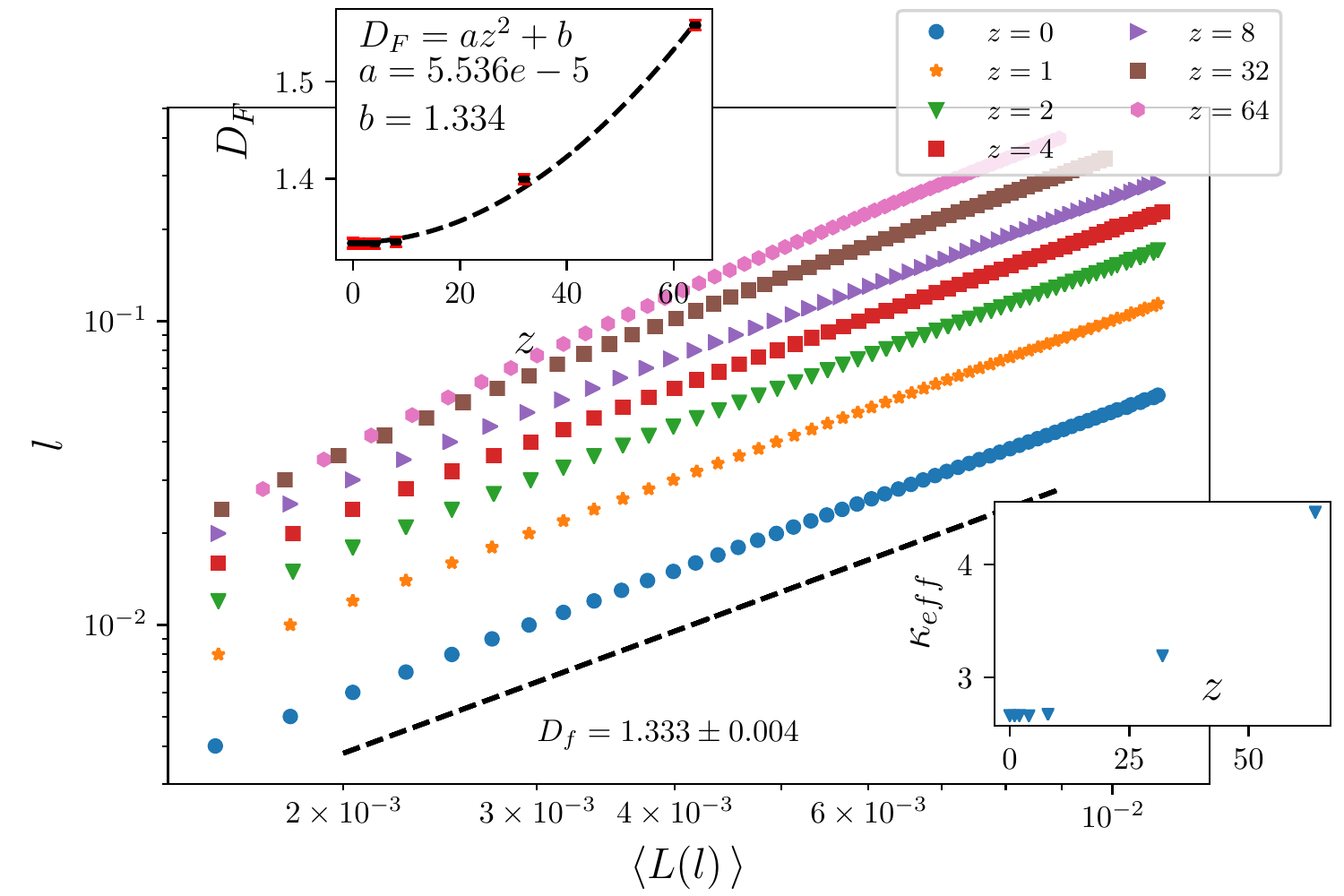}
		\caption{}
		\label{fig:D_f_k=8_3}
	\end{subfigure}
	\begin{subfigure}{0.45\textwidth}\includegraphics[width=\textwidth]{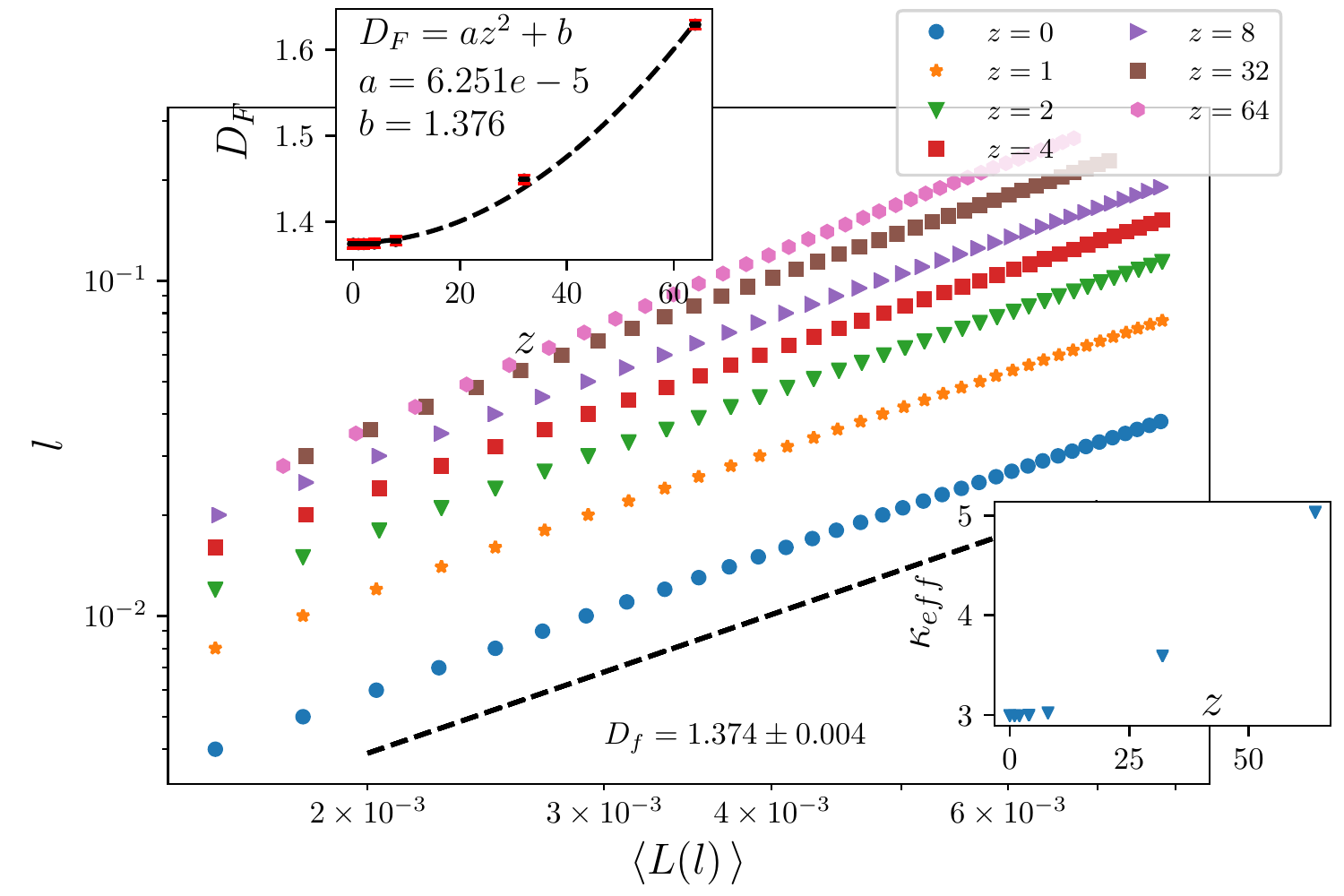}
		\caption{}
		\label{fig:D_f_k=3}
	\end{subfigure}
	\begin{subfigure}{0.45\textwidth}\includegraphics[width=\textwidth]{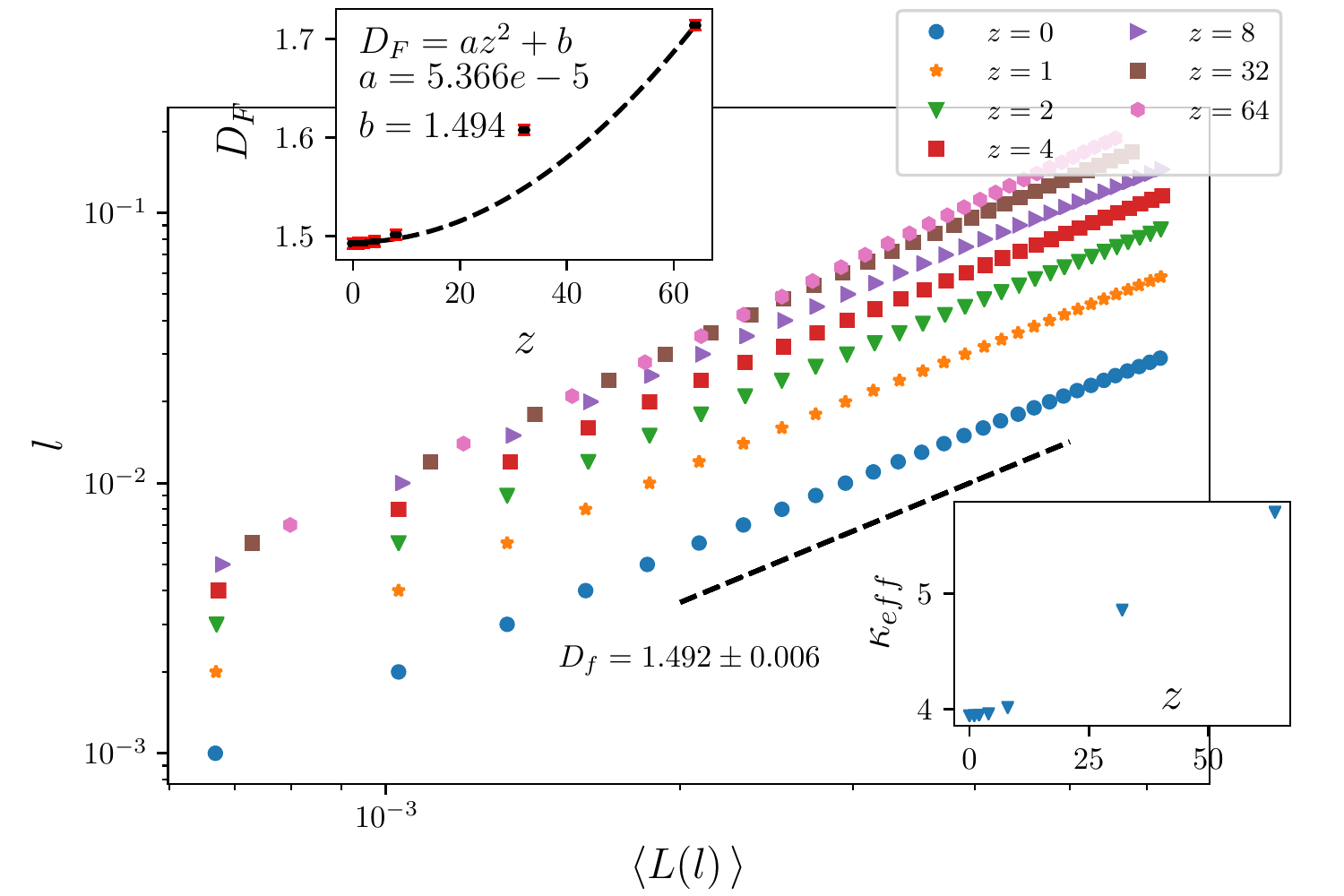}
		\caption{}
		\label{fig:D_f_k=4}
	\end{subfigure}
		\caption{(Color online) The log-log plot of $l-L$ for the initial diffusivity parameter (a) $\kappa=2$, (b) $\kappa=\frac{8}{3}$, (b) $\kappa=3$ and (d) $\kappa=4$. Upper insets: The calculated fractal dimension $D_f$ in terms of $z$. Lower insets: The calculated $\kappa_{\text{effectice}}\equiv 8(D_f-1)$.}
	\label{fig:DF}
\end{figure*}

Now let us go to the calculations of the mixed model. We should do the calculations for all $\kappa$s separately to observe how the statistics depends on the initial $\kappa$. This has been done in Fig.~\ref{fig:DF} for which the slopes of the log-log plot are the fractal dimension of the curves. We have found that the relation between the loop lengths $l$ and the linear size of the trace $L$ is power-law with non-trivial exponent, which depends on the $z$ value. The initial $\kappa$ values are $\kappa=2$ (Fig.~\ref{fig:D_f_k=2}), $\kappa=\frac{8}{3}$ (Fig.~\ref{fig:D_f_k=8_3}), $\kappa=3$ (Fig.~\ref{fig:D_f_k=3}) and $\kappa=4$ (Fig.~\ref{fig:D_f_k=4}), which correspond to $c=-2$, $c=0$, $c=\frac{1}{2}$ and $c=1$ conformal field theories respectively. This fractal dimension grows with second power of $z$ for all values of $\kappa$, i.e. $D_F(z)-D_F(z=0)\sim z^2$. This has been shown in the insets of all graphs, in which the proportionality constants have been shown. The numerical values of $D_F(z=0)$ ($b$ in the graphs) are also well known, i.e. $D_F^{\kappa=2}(z=0)=\frac{5}{4}$, $D_F^{\kappa=8/3}(z=0)=\frac{4}{3}$, $D_F^{\kappa=3}(z=0)=\frac{11}{8}$ and $D_F^{\kappa=4}(z=0)=\frac{3}{2}$. The above analysis show that the presence of these type of random scatterers change considerably the properties of SLE traces. The increasing of the fractal dimension shows that the traces become more compact and twisted, which is reasonable having the localization effects of the disordered media in mind. Such a classical disorder-induced localization has also been seen in many other systems, ranging from two-dimensional electron gas~\cite{Najafi2018percolation} to quantum Hall effect~\cite{Girivin}.\\
\begin{figure*}
	\centering
	\begin{subfigure}{0.40\textwidth}\includegraphics[width=\textwidth]{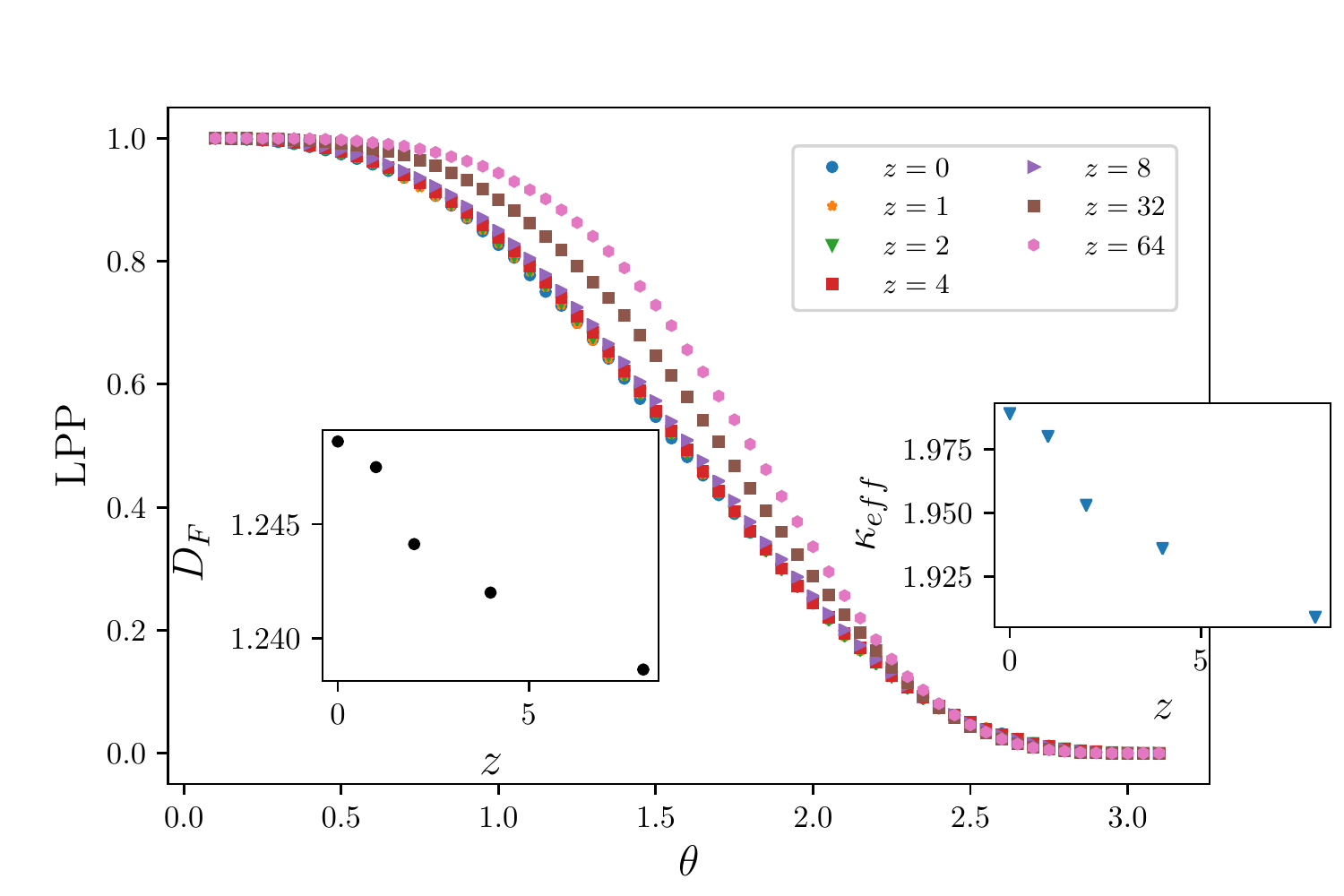}
		\caption{}
		\label{fig:k=2lpp}
	\end{subfigure}
	\begin{subfigure}{0.45\textwidth}\includegraphics[width=\textwidth]{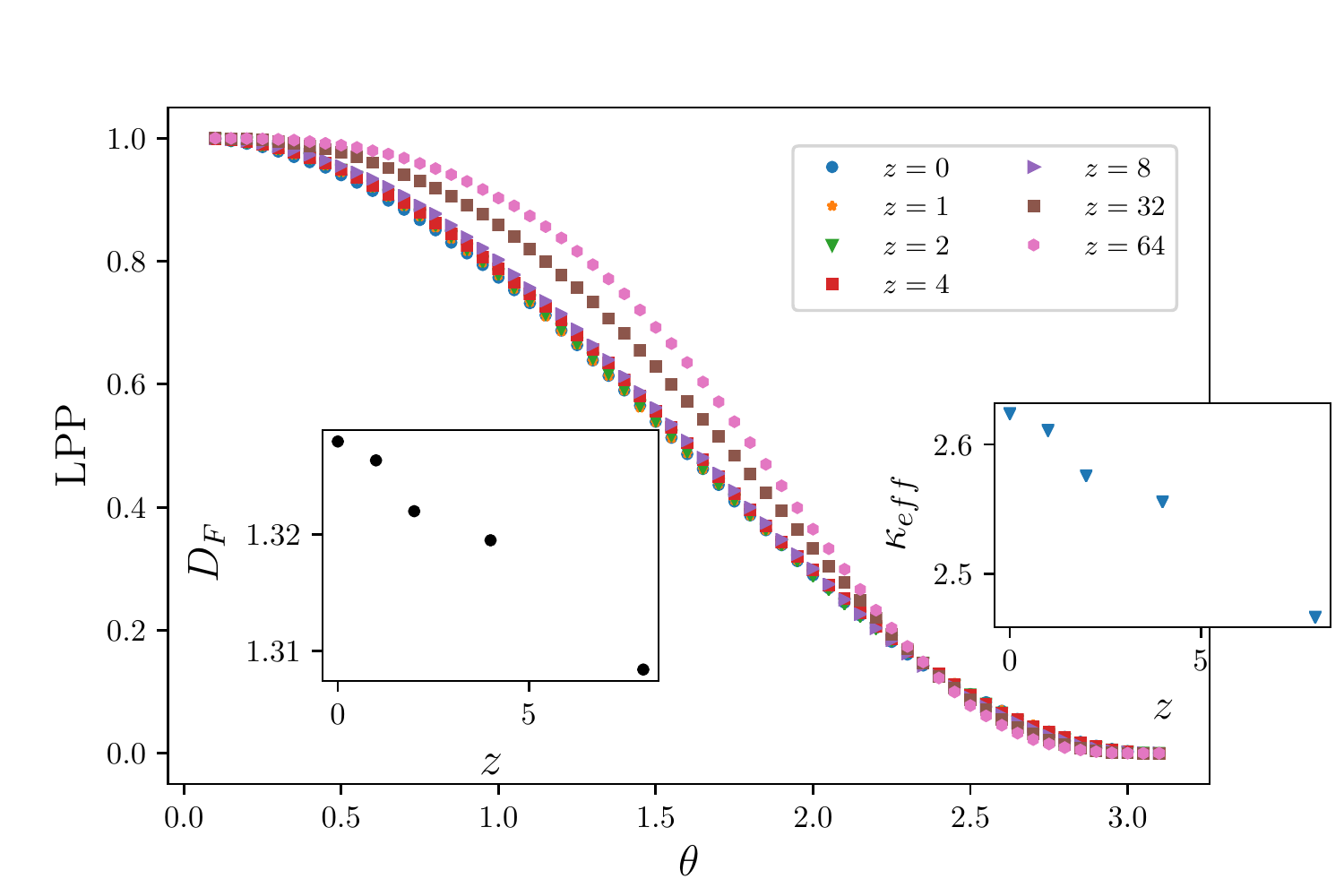}
		\caption{}
		\label{fig:k=8_3lpp}
	\end{subfigure}
	\begin{subfigure}{0.45\textwidth}\includegraphics[width=\textwidth]{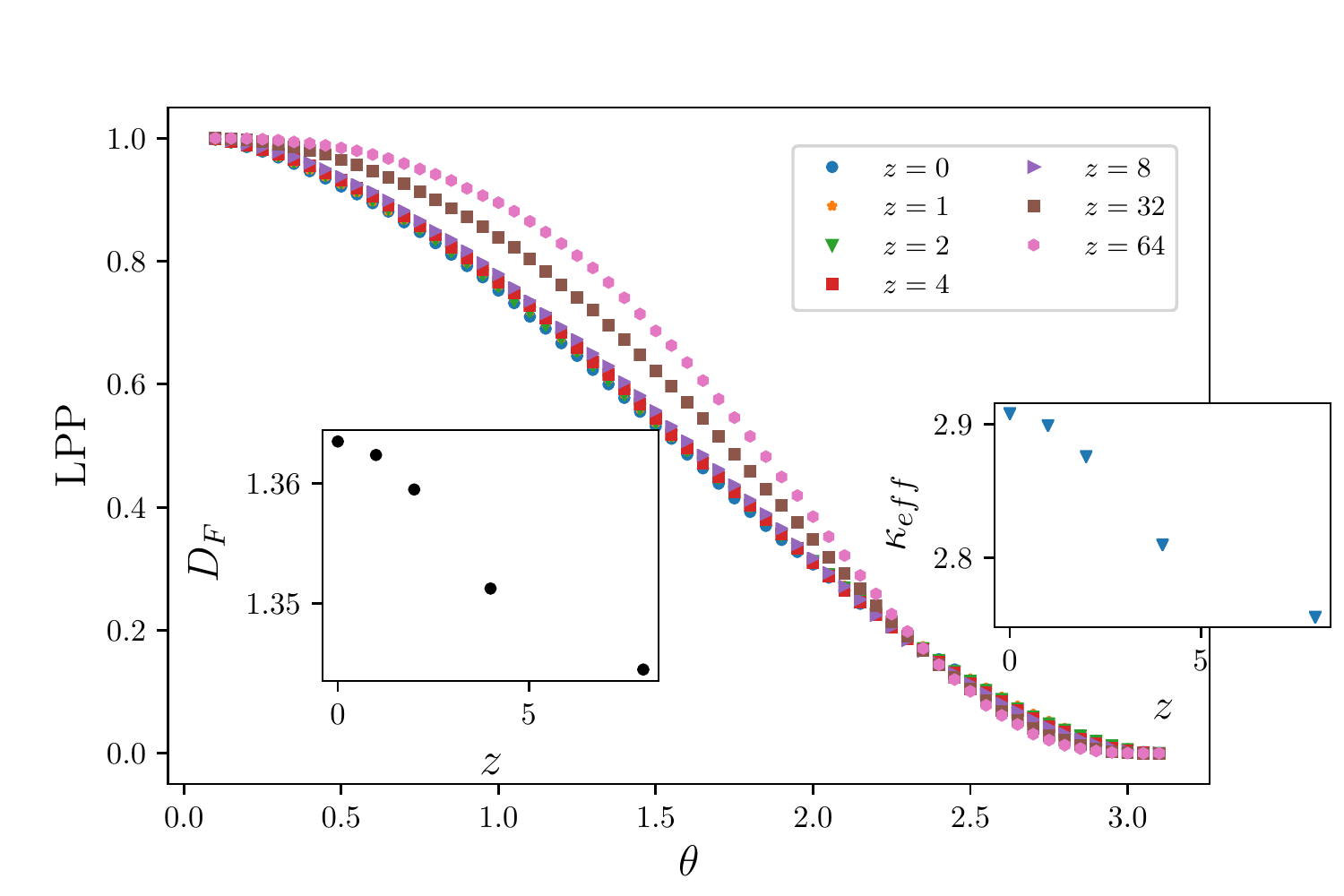}
		\caption{}
		\label{fig:k=3lpp}
	\end{subfigure}
	\begin{subfigure}{0.45\textwidth}\includegraphics[width=\textwidth]{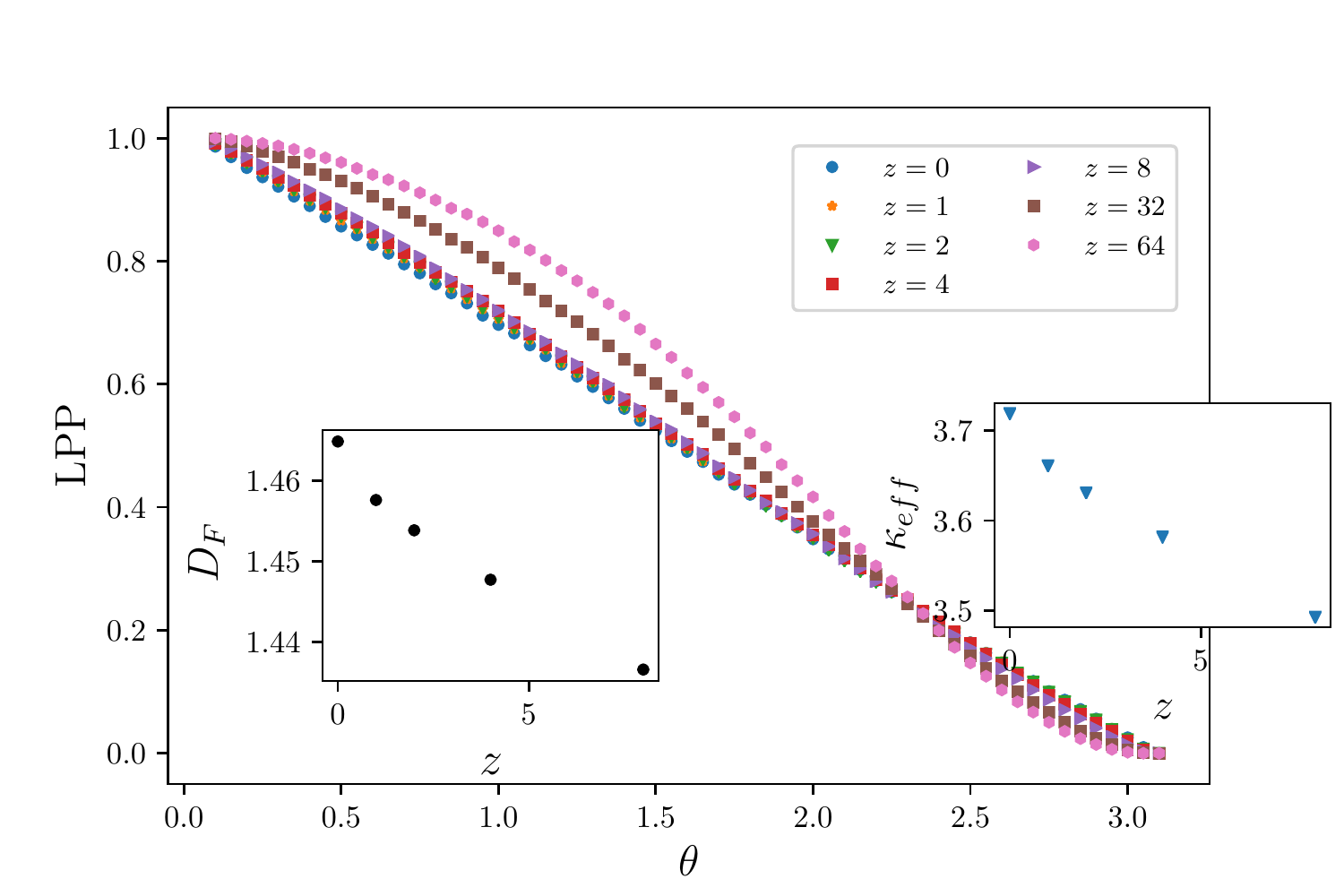}
		\caption{}
		\label{fig:k=4lpp}
	\end{subfigure}
	\caption{(Color online) The LPP$\theta$ for the initial diffusivity parameter (a) $\kappa=2$, (b) $\kappa=\frac{8}{3}$, (c) $\kappa=3$ and (d) $\kappa=4$. Left insets: $\kappa$ obtained by fitting of the plots, in terms of $z$. Right insets: The corresponding fractal dimension $D_f^{\text{effective}}\equiv 1+\kappa/8$.}
	\label{fig:LPP}
\end{figure*}
\\
The scaling behavior of $l-L$ graphs does not imply the conformal symmetry, and some other tests are needed. As stated above, we have calculated the left passage probability, which has been shown in Fig.~\ref{fig:LPP}. The graphs have been fitted by means of Eq.~\ref{Eq:LPP}, form which the effective diffusvity can be extracted. We see from the graphs that not only $D_F$ and $\kappa_{\text{effective}}$ decrease with $z$, but for large $z$ values, the fitting of the graphs by Eq.~\ref{Eq:LPP} becomes ill-defined, since the behavior of the graphs change considerably and show vast deviations from that which is theoretically predicted. For small $z$ however, although the fitting is good and reliable, we see that the fractal dimension shows a decreasing behavior with $z$ which is in contrast to the result obtained from the scaling behavior of $l-L$ graphs.\\
As mentioned in the previous sections, for conformal invariant systems one expects that the fractal dimension of the interfaces is related to $\kappa$ by $D_f=1+\kappa/8$, and increasing in the $\kappa$ value results to increasing of the fractal dimension. Therefore, these effects show that the system in non-zero $z$ fails to be conformal invariant.

\section{conlusion}
In this paper we have considered the effect of the manipulating of the space through which the SLE trace grows. We introduced the Henon-percolation lattice. The Henon map is appropriate for this purpose, since it is a two-dimensional map involving unstable and stable fixed points, which play the role of repulsive and attractive points in the lattice respectively. The points of the lattice have been randomly chosen to be repulsive or attractive. We have presented some theoretical equations which couple the SLE theory with the Henon dynamics by growing the SLE traces in the background of Henon-percolation system. We have controlled this coupling by the $z$ parameter, which favors the Henon dynamics, i.e. $z=0$ is the regular SLE.\\
The dilute phase of SLE was investigated, i.e. $\kappa=2,\frac{8}{3},3$ and $4$, and their behavior under increasing the amount of $z$ has been investigated. Two kind of SLE tests were presented: the fractal dimension of the traces $D_f$ (which was obtained by box-counting method), and left passage probability LPP($\theta$). We obtained that the fractal dimension of the curves grow with second power of $z$, i.e. $D_f(z)-D_f(z=0)\sim z^2$. This is in contrast to LPP analysis which show a decreasing behavior in terms of $z$. We explicitly show that the conformal symmetry breaks down for non-zero $z$, and the traces are conformal only for $z=0$ for which the relation $D_f=1+\kappa/8$ holds. \\
In many situations we have random traces (which are supposed to be SLE traces in the regular host system) in the background of random defects/scatterers. The above analysis show that the presence of these random scatterers (when are modeled by the Henon map) change considerably their properties. The example is the increasing of the fractal dimension, meaning that the traces become more twisted, which is reasonable having the localization effects of the disordered media mind. Additionally, the absence of consistency between $D_f$ and LPP show that the conformal invariance is lost, i.e. the traces are no longer SLE traces for non-zero $z$s.

\end{document}